\documentclass[twocolumn,amsmath,amssymb,aps,pra]{revtex4-2}
\usepackage{bm}
\usepackage{amsfonts}
\usepackage{amsmath}
\usepackage{graphicx}
\usepackage{float}

\begin{document} 
\title{Properties of ``special'' hyperbolic Bessel-Gaussian optical beams}
\author{Tomasz Rado\.zycki}
\email{t.radozycki@uksw.edu.pl}
\affiliation{Faculty of Mathematics and Natural Sciences, College of Sciences, Institute of Physical Sciences, Cardinal Stefan Wyszy\'nski University, W\'oycickiego 1/3, 01-938 Warsaw, Poland} 

\begin{abstract}
An explicit formula for a new type of beams, which in this work are called the "special" hyperbolic Bessel-Gaussian  (SHBG) beams, has been derived, using the method of the Hankel transform formulated in our previous work. The fundamental properties of these beams are analyzed. The parameters that define the beam shape have been identified and related to those of the fundamental Gaussian beam. The analytical expressions for the SHBG beams include an additional parameter $\gamma$, which allows the beam's shape to be modified to some extent. In the plane perpendicular to the propagation direction, these beams exhibit the annular nature. Interestingly, initially (i.e. near the beam's spot) a single ring splits into a number of rings as one is moving along the beam.   This is especially apparent for $\gamma$ close to unity, as this effect then appears for values of $z$ relatively small compared to the Rayleigh length i.e., where the energy concentration in the beam is still high. The phase of the wave, whose behavior is in certain aspects typical of modes having the vortex character, is also studied in this paper. 
\end{abstract}

\maketitle

\section{Introduction}\label{intr}

One of the most growing research areas in optics is that involving the description, generation and practical applications of various laser beams. Among these, Gaussian-type beams, especially those of a cylindrical nature, are notable as solutions to the paraxial wave equation. Their applications cover a very wide area. To enumerate only the most important ones, we will mention here optical trapping and guiding of particles, image processing, optical communication, harmonics generation in nonlinear optics, quantum cryptography and even biology and medicine~\cite{ste,fazal,pad,woe,bowpa,grier1,kol,alt,nis,sas,cc,trknot}. The literature on this subject is extremely vast, so by necessity a subjective choice needs to be made. The most important example are pure Gaussian beams~\cite{davis,sie,nemo,mw,saleh,ibbz,sesh,gustavo,er,selina} which can be endowed with orbital angular momentum (OAM). Somewhat more complex in nature are Bessel-Gaussian (BG)~\cite{saleh,gori,april1,mendoza} and Laguerre-Gaussian (LG) beams~\cite{sie,saleh,mendoza,lg,lg2,april2,april3,nas}, which exhibit a ring-like pattern of the irradiance in the transverse plane and possess a vortex core. 

These mentioned examples are of great importance since for them rigorous analytical expressions exist (naturally within the mentioned paraxial approximation). Moreover, their nontrivial structure, which is very far from idealized and purely textbook case of plane waves, can lead to various intriguing phenomena. Any further solution of this type that might be found seems valuable and potentially useful. Designing and experimentally generating an increasing variety of light beams offers the possibility of shaping them to better suit experimental requirements or practical applications. 
 
In the present work we would like to describe a new family of beams, which are exact solutions of the paraxial Helmholtz equation, and analyze their basic properties. They are, like the previously mentioned beams, eigenstates of OAM and as such have vortex nature (for nonzero OAM). Throughout the paper they will be provisionally called  ``special'' hyperbolic Bessel-Gaussian (SHBG) beams. Due to the Gaussian fall-off (albeit slightly distinct from the aforesaid beams), these modes are square integrable in the plane transverse to the direction of propagation and therefore they carry a finite amount of energy so they might be experimentally realized. The additional parameter $\gamma$ which appears in the solution of the paraxial equation enables, to some extent, the shape of the beam to be modified and tailored to one's expectations.

The electric field of a light wave propagating along the $z$-axis, within the standard scalar approximation 
($\bm {\nabla}(\bm{\nabla E})\longmapsto 0$),
can be represented through the scalar envelope $\Psi(\bm{r},z,t)$ and a constant vector $\bm{E}_0$ in the form
\begin{equation}
\bm{E}(\bm{r},z,t)=\bm{E}_0\Psi(\bm{r},z,t),
\label{ep}
\end{equation}
with $\bm{r}$ denoting transverse coordinates only. For many applications this assumption is acceptable, although ultimately the full vector beams shall be of interest. In what follows, however, we focus on the properties of the function $\Psi$, which satisfies the d'Alembert equation, without regard to any polarization effects. 

In the paraxial regime the d'Alembert equation can be simplified by first substituting
\begin{equation}
\Psi(\bm{r},z,t)=e^{ik(ct-z)}\psi(\bm{r},z),
\label{epi}
\end{equation}
and next, under the assumption of $\psi(\bm{r},z)$ being a slowly varying function of $z$,  neglecting the second order derivative with respect to $z$ thanks to the common assumption
\begin{equation}
|\partial^2_z\psi|\ll |k\partial_z\psi|.
\label{parap}
\end{equation}  
The symbol $\partial_z$ stands here for the partial derivative $\frac{\partial}{\partial z}$. This simplification, without going into the details that can be found elsewhere~\cite{sie,lax}, leads to the well-known equation, called the {\em paraxial equation}, for the envelope:
\begin{equation}\label{paraxial}
\mathcal{4}_\perp \psi(\bm{r},z)-2ik\partial_z \psi({\bm{r}},z)=0,
\end{equation}
where the Laplace operator $\mathcal{4}_\perp$ stands for two-dimensional one acting in the transverse plane only. 

Since we are concerned with cylindrical-type beams it is convenient to switch to the polar coordinates $r=\sqrt{x^2+y^2}$ and $\varphi$ and isolate the dependence on the angular variable by means of the substitution
\begin{equation}
\psi(r,\varphi,z)=e^{i n\varphi}\Phi(r,z),\;\;\;\mathrm{where}\,\;\; n\in\mathbb{Z}.
\label{phi}
\end{equation}

This yields the paraxial equation in the required form
\begin{equation}\label{parcylf}
\Big(\partial_r^2+\frac{1}{r}\,\partial_r-\frac{n^2}{r^2}-2ik\partial_z\Big)\Phi(r,z)=0.
\end{equation}
which provides the starting point for deriving expressions for various axial beams.  As mentioned, in this paper we will be interested in a peculiar beam -- a ``special'' hyperbolic Bessel-Gaussian beam -- which seems not to have been analysed in the literature so far.  

In the next section analytical formulas for the shape of this beam will be derived, employing the method based on the Hankel transform, developed in our former paper. In Section~\ref{bp} we will study the parameters characterizing the beam, and relate them to those occurring in the fundamental beam of cylindrical type, i.e., the Gaussian beam. In Section~\ref{pc}, the simplest example from the considered class will be presented, with a specific choice of parameter values. Section~\ref{int} deals with the transverse intensity distribution of the beam and deviations from the commonly employed beams will be highlighted. In the last section we address the phase of the wave and study its behavior both in the transverse plane and along the propagation axis.

\section{Derivation of the ``special'' hyperbolic Bessel-Gaussian beam}\label{der}

In the previous paper~\cite{trhan} a concise method based on the use of the Hankel transform, applicable for straightforward derivation of various cylindrical beams, was formulated. Below it will be exploited to generate an expression for a SHBG beam. This is a new type of a paraxial beam with some peculiar characteristics that -- up to our knowledge -- has not been discussed in the literature. The word ``special'' is used to highlight the contrast with ``ordinary'' hyperbolic Bessel-Gaussian beam: the argument of the Bessel function is now quadratic (i.e. proportional to $r^2$), whereas its index is halved.

The method formulated in~\cite{trhan} relies upon applying to Eq.~(\ref{parcylf}) the $n$th-order Hankel transform with respect to the radial variable $r$, which for a given function $f(r)$ is defined as~\cite{patra}:
\begin{equation}
\hat{f}(s)=\int\limits_0^\infty \mathrm{d}r\, r J_n(sr)f(r),
\label{hankel}
\end{equation}
where $J_n(x)$ is the Bessel function of the first kind. The paraxial equation is then simplified to the elementary first-order differential equation:
\begin{equation}
(2ik\partial_z+s^2)\hat{\Phi}_n(s,z)=0,
\label{phit1}
\end{equation}
which can be immediately solved in the trivial way leading to
\begin{equation}
\hat{\Phi}_n(s,z)=g_n(s)e^{is^2z/2k}.
\label{phit2}
\end{equation}

It should be pointed out that the function $g_n(s)$ bearing the index $n$ not necessarily depends on $n$. Its form can be ``freely'' chosen, with the obvious restriction that the inverse transform~(\ref{hankelinv}) is to exist. The role of the index $n$ is merely confined to remind that this inverse transform leading to the explicit form of $\Phi(r,z)$ has to be of exactly $n$th order. It is defined through the identical integral as in~(\ref{hankel})~\cite{patra}, i.e.,
\begin{equation}
f(r)=\int\limits_0^\infty \mathrm{d}s\, s J_n(r s)\hat{f}_n(s).
\label{hankelinv}
\end{equation}
which yields the solution of Eq.~(\ref{paraxial}) in the form
\begin{equation}
\psi(r,\varphi,z)=e^{in\varphi}\int\limits_0^\infty \mathrm{d}s\, s J_n(r s )g_n(s)e^{is^2z/2k}.
\label{inp}
\end{equation}
This type of an equation has already appeared in the context of conical refraction~\cite{sok,turp,myl}.

Let us now pick the function $g_n(s)$ as follows: 
\begin{equation}
g_n(s)=I_{\frac{n}{2}}(\chi s^2/4)e^{-w_0^2s^2/4},
\label{gfun}
\end{equation}
where $I_{\frac{n}{2}}$ represents the hyperbolic Bessel function of the order $\frac{n}{2}$, $\chi$ is a certain constant to be chosen later, and $w_0$ will turn out to be closely related to the beam's waist (in Sec.~\ref{bp} a distinction will be made between the value of beam's waist for a Gaussian beam and for the beam considered in this work). This form fulfills the condition of integrability of~(\ref{inp}), and the presence of the Gaussian factor $e^{-w_0^2s^2/4}$ ensures that the transverse intensity profile of the resulting beam will be of similar character.
Consequently the envelope for the beam in question can be written as
\begin{eqnarray}
\Psi_{nk}&&(r,\varphi,z,t)=e^{ik(ct-z)}\psi(r,\varphi,z)=\label{inpf}\\
&&e^{ik(ct-z)}e^{in\varphi}\!\int\limits_0^\infty\! \mathrm{d}s\, s J_n(r s )I_{\frac{n}{2}}(\chi s^2/4)e^{-\alpha(z)s^2/4},
\nonumber
\end{eqnarray}
where $\alpha(z)=w_0^2 -2iz/k$. Performing the integration with respect to $s$ one comes to the final result:
\begin{eqnarray}
\Psi_{nk}(r,\varphi,z,t)=&&\;N_n\, e^{ik(ct-z)}e^{in\varphi}e^{-\frac{\alpha(z)r^2}{\alpha^2(z)-\chi^2}}\label{qhbg}\\
&&\;\times\frac{2}{\sqrt{\alpha^2(z)-\chi^2}} \;I_{\frac{n}{2}}\Big(\frac{\chi\, r^2}{\alpha^2(z)-\chi^2}\Big),
\nonumber
\end{eqnarray}
with $N_n$ standing for a certain normalization constant. For the inverse transform~(\ref{inpf}) to exist, it has been assumed that $\chi<w_0^2$. It will be then convenient to introduce a dimensionless parameter $\gamma$ and write
\begin{equation}
\chi=\gamma w_0^2,\;\;\;\; \mathrm{with}\;\; 0<\gamma<1.
\label{gamma}
\end{equation}
The most interesting case emerges when $\gamma$ is approaching unity. At the other end of the interval, i.e., for $\gamma\rightarrow 0$ the hyperbolic Bessel-Gaussian beam is recovered, to which we will come back later.

Substituting the expression obtained in~(\ref{qhbg}) into~(\ref{paraxial}) it can be verified in a straightforward way that it represents a true paraxial beam regardless of the values of the parameters $\chi$ (i.e., $\gamma$) and $w_0$.

As it was shown in~\cite{trhan} expression~(\ref{qhbg}) can represent the solution of the d'Alembert equation as well, upon the substitution $\alpha(z)\mapsto \alpha((z+ct)/2)$. In this case, however, the beam (or rather pulse) would not be monochromatic, so that $k$ and $\omega$ would no longer be related by the condition $\omega= c k$.

Due to the presence of the factor $e^{i n\varphi}$ the modes described with the formula~(\ref{qhbg}), carrying different values of the orbital angular momentum, are orthogonal to each other. In order to establish the value of the constant $N_n$ it is required that in the perpendicular plane the following condition holds:
\begin{equation}
\int \mathrm{d}^2r |\Psi_{nk}(r,\varphi,z,t)|^2=1.
\label{norm}
\end{equation}
This condition is only an approximate one since for large $r$ the paraxial approximation is not justified. However, this domain contributes very little to the integral. The most easily is to calculate this quantity for $z=0$. Introducing a new integration variable $u=2 r^2/w_0^2(1-\gamma^2)$ and using the integral:
\begin{eqnarray}
\int\limits_0^\infty \mathrm{d}u e^{-u}I_\nu^2(a u)=&&\,\frac{4^\nu a^{2\nu}\Gamma^2(1/2+\nu)}{\pi \Gamma(2\nu+1)}\label{norm1}\\
&&\,\times \,{}_2F_1\Big(\nu+\frac{1}{2},\nu+\frac{1}{2};2\nu +1, 4a^2\Big),
\nonumber
\end{eqnarray}
where ${}_2F_1$ stands for the hypergeometric function and $a<1/2$ is a constant parameter, it can be shown that the normalization coefficient takes the value of
\begin{equation}
|N_n|=\frac{w_0}{\Gamma\big(\frac{n+1}{2}\big)}\, \left[\frac{n!}{2\gamma^n\,{}_2F_1\big(\frac{n+1}{2},\frac{n+1}{2};n+1;\gamma^2\big)}\right]^{1/2}.
\label{Nvalue}
\end{equation}
As it is seen for $\gamma<1$ the condition of the integrability of~(\ref{norm1}) is met. It should be also noted that for small $\gamma$ this expression becomes proportional to $\gamma^{-n/2}$. This fact will be made use of in Sect.~\ref{pc}

\section{Beam's parameters}\label{bp}

From now on we have to differentiate between parameters characterizing an ordinary Gaussian beam, which will serve as our standard reference, and those related to SHBG beam. For this reason the letter ``G'' will be added as a lower index of the parameter designations related to the former: the symbols $w_{G0}, w_G(z), z_{GR}, R_G(z), \varUpsilon_G(z)$ will, respectively, refer to the waist, width, Rayleigh length, radius of the wavefront curvature and Gouy phase of the Gaussian beam and identically but without letter ``G'' in the case of the SHBG beam. 

When inspecting the form of the standard Gaussian beam~\cite{saleh} it is clear that the analog of the complex $q$-parameter  (i.e., $q(z)=z+iz_{GR}$) may be now read off the formula~(\ref{qhbg}). If one temporarily ignores the Bessel function $I_{\frac{n}{2}}$ and merely accounts for the explicit Gaussian factor, the role of the parameter $q(z)$  seems to be now played by the quantity
\begin{eqnarray}
q(z)\;\;&&\leadsto\;\;\frac{ik}{2}\left(\alpha(z)-\frac{\chi^2}{\alpha(z)}\right),\label{anq}\\
&&\mathrm{where}\;\;\;\; \alpha(z)=w_{G0}^2\left(1-i\frac{z}{z_{GR}}\right).\nonumber
\end{eqnarray}
The standard analysis~\cite{saleh} would then lead to the following values of the fundamental parameters:
\begin{subequations}\label{paramn}
\begin{align}
&w_0=w_{G0}\sqrt{1-\gamma^2},\label{paramn1}\\
&w(z)=w_{G0}\sqrt{\frac{2}{(1+\gamma)h_+(z)+(1-\gamma)h_-(z)}}, \label{paramn2}\\
&R(z)=\frac{z_{GR}^2}{z}\cdot\frac{2}{h_+(z)+h_-(z)}, \label{paramn3}\\
&z_R=z_{GR}\sqrt{1-\gamma^2}, \label{paramn4}
\end{align}
\end{subequations}
where the $\chi$ parameter has been replaced with $\gamma w_0^2$ according to~(\ref{gamma}) and simultaneously two auxiliary functions have been introduced:
\begin{equation}
h_\pm(z)=\frac{1}{(1\pm\gamma)^2+z^2/z_{GR}^2},
\label{hpm}
\end{equation}
in order to avoid lengthy expressions.
However, these formulas would lead to a conclusion which stays at odds with the intuition stemming from the uncertainty principle, regarding a number quantifying the beam quality~\cite{saleh}, namely
\begin{equation}
{\mathbb M}^2=\frac{\pi w_0 \theta_0}{\lambda}=\sqrt{1-\gamma^2}<1.
\label{m2n}
\end{equation}
The aperture half-angle $\theta_0$ which appears above is defined as the limit of the ratio between the beam's width and the distance:
\begin{equation}
\theta_{0}\approx \tan\theta_0=\lim_{z\rightarrow\infty}\frac{w(z)}{z}
\label{theta0}
\end{equation}
This result would effectively mean that by choosing $\gamma\lesssim 1$ one could make the beam arbitrarily good. It is, therefore, obvious that in our considerations the Bessel factor, which modifies the Gaussian behavior at large distances, needs to be accounted for as well.

In order to recognize the actual behavior of $\Psi_{nk}(r,\varphi,z,t)$ in the perpendicular plane for large radial variable $r$, it is necessary to invoke the corresponding expansion of the function $I_\nu(\xi)$ as $|\xi|\gg 1$~\cite{span}:
\begin{equation}
I_\nu(\xi)\sim \frac{e^\xi}{\sqrt{2\pi \xi}}\left(1-\frac{1/4-\nu^2}{2 \xi}+\ldots\right).
\label{appri}
\end{equation}
Applying this formula, one can write
\begin{equation}
\Psi_{nk}(r,\varphi,z,t)\sim N_n \sqrt{\frac{2}{\pi\chi}}\,\frac{1}{r}\,e^{ik(ct-z)}e^{in\varphi-in\pi/4}e^{-\frac{r^2}{\alpha(z)+\chi}},
\label{psiapp}
\end{equation}
and observe that now, instead of~(\ref{anq}), there is a correspondence
\begin{equation}
q(z)\;\leadsto\;\;\frac{ik}{2}(\alpha(z)+\chi)=z+i(1+\gamma)z_{GR}.\label{aq}
\end{equation}
Consequently, following the standard procedure, one derives the set of parameters in the form
\begin{subequations}\label{parama}
\begin{align}
&w_0=w(0)=w_{G0}\sqrt{1+\gamma},\label{parama1}\\
&w(z)=w_{G0}\sqrt{1+\gamma}\left(1+\frac{z^2}{(1+\gamma)^2z_{GR}^2}\right)^{1/2}, \label{parama2}\\
&R(z)=z\left(1+(1+\gamma)^2\frac{z_{GR}^2}{z^2}\right). \label{parama3}
\end{align}
\end{subequations}
These expressions can be further simplified if the Rayleigh length $z_R$ appropriate to the considered SHBG beam is introduced instead of $z_{GR}$. This new length is defined in the standard way, through the requirement
\begin{equation}
\frac{w(0)^2}{w(z_R)^2}=\frac{1}{2},
\label{rldef}
\end{equation}
which entails
\begin{equation}
z_R=(1+\gamma)z_{GR}.
\label{rl}
\end{equation}
Now, the $q$-parameter acquires its standard definition (i.e., $q(z)=z+iz_{R}$) and the set of relations~(\ref{parama}) can be given the form
\begin{subequations}\label{param}
\begin{align}
&w(z)=w_{0}\sqrt{1+\frac{z^2}{z_R^2}}, \label{param1}\\
&R(z)=z\left(1+\frac{z_R^2}{z^2}\right), \label{param2}\\
&\varUpsilon(z)=\frac{n+1}{2}\arctan\left(\frac{2z/z_R}{1-\gamma-(1+\gamma)z^2/z_R^2}\right). \label{param3}
\end{align}
\end{subequations}
As can be seen, the wavefront-curvature radius is defined by the same formula as for a Gaussian beam. However, it must be remembered that the formula~(\ref{param2}) holds only for large values of $r$. Closer to the propagation axis, some modifications originating from the Bessel function $I_{\frac{n}{2}}$ appear, as will be illustrated in the following sections. For extreme values of $z$, one obtains the usual results:
\begin{equation}
\infty\underset{z\rightarrow 0}{\longleftarrow}\;\;\; R(z)\;\;\underset{z\rightarrow \infty}{\longrightarrow}\infty,\label{gr1}
\end{equation}
and, in turn, the minimal value is obtained at the Rayleigh length: $R(z_R)=2z_R$, which is again the standard result. 

Finally, it is nice to observe that 
\begin{equation}
\theta_0=\frac{\theta_{G0}}{\sqrt{1+\gamma}}
\label{to}
\end{equation}
and the SHBG beam turns out to be equally good, as regards the transverse spread, as the Gaussian beam:
\begin{equation}
{\mathbb M}^2=\frac{\pi w_0 \theta_0}{\lambda}=1.
\label{m2}
\end{equation}

\section{Particular cases}\label{pc}

Beams of the discussed sort assume an especially simplified form for half-integer values of the index of Bessel functions. If one sets $n=2l+1$, with $l$ being an integer, they get reduced to plain sums of merely few terms, thanks to the mathematical formula~\cite{gr}:
\begin{eqnarray}
&&I_{l+1/2}(\xi)=\frac{1}{\sqrt{2\pi \xi}}\label{half}\\
&&\;\;\;\;\times\sum_{m=0}^l\frac{(m+l)!}{m!(l-m)!(2\xi)^m}\left[(-1)^me^\xi-(-1)^le^{-\xi}\right].
\nonumber
\end{eqnarray}
Note that in this case, regardless of the value of $l$ and the number of terms in the sum, only two Gaussian factors survive in this expression. These are, when combined with the analogous factor in formula~(\ref{qhbg}):
\begin{equation}
\exp\bigg(\frac{-r^2}{w_0^2(1-i\frac{z}{z_R})}\bigg)\;\;\;\;\mathrm{and}\;\;\;\; \exp\bigg(\frac{-r^2}{w_0^2(\frac{1-\gamma}{1+\gamma}-i\frac{z}{z_R})}\bigg).
\label{gafa}
\end{equation}
Of these, in the interesting case of $\gamma$ approaching unity, the former becomes dominant for larger values of $r$ (this was already seen in expression~(\ref{psiapp})). On the other hand, close to the $z$-axis both are necessary since, otherwise, unphysical divergences would be observed. 

For $l=0$ a further drop of the number of terms occurs and there remain two terms only (the overall constant phase is henceforth omitted):
\begin{eqnarray}
&&\Psi_{1k}(r,\varphi,z,t)=\frac{1}{\sqrt{-\pi\ln (1-\gamma^2)}}\,e^{ik(ct-z)}e^{i\varphi}\label{L0}\\
&&\;\;\;\;\times\frac{1}{r}\left[\exp\bigg(\frac{-r^2}{w_0^2(1-i\frac{z}{z_R})}\bigg)-\exp\bigg(\frac{-r^2}{w_0^2(\frac{1-\gamma}{1+\gamma}-i\frac{z}{z_R})}\bigg)\right],
\nonumber
\end{eqnarray}
Each of these two terms satisfies the paraxial equation~(\ref{paraxial}), but they are divergent as $r\rightarrow 0$. Due to the cancellations, this does not happen if both terms are kept together (likewise $l+1$ expressions in formula~(\ref{half})). It also follows from the application of the formula~(\ref{lims}) below to the case $r\ll 1$ that the beam of index $n$, independently of the value of $\gamma$, displays a vortex of order $n$ along the axis of propagation, identically to the $n$th-order Gaussian, Bessel-Gaussian or Laguerre-Gaussian beams.

The other special case arises when $\gamma\rightarrow 0$. Due to the known limit~\cite{gr} 
\begin{equation}
\xi^{-n/2}I_{\frac{n}{2}}(a\, \xi)\;\longrightarrow\; \frac{\big(\frac{a}{2}\big)^{n/2}}{\Gamma\big(1+\frac{n}{2}\big)} 
\label{lims}
\end{equation} 
as $\xi\rightarrow 0$, it is straightforward to show that one obtains the Gaussian beam of order $n$. This is obvious when looking at the integral~(\ref{inpf}) (with the normalization coefficient~(\ref{Nvalue}) which contains the factor $\gamma^{-n/2}$). It is also clear when inspecting the expression~(\ref{gfun}). Taking the limit $\chi\rightarrow 0$ (i.e. $\gamma\rightarrow 0$) makes the function $g_n(s)$ acquire the form
\begin{equation}
g_n(s)=s^ne^{-w_{G0}^2s^2/4},
\label{gfun1}
\end{equation}
which constitutes the generating function for the Gaussian beam of order $n$~\cite{trhan}.

\section{Intensity profile}\label{int}

\begin{figure}[h]
\begin{center}
\includegraphics[width=0.50\textwidth,angle=0]{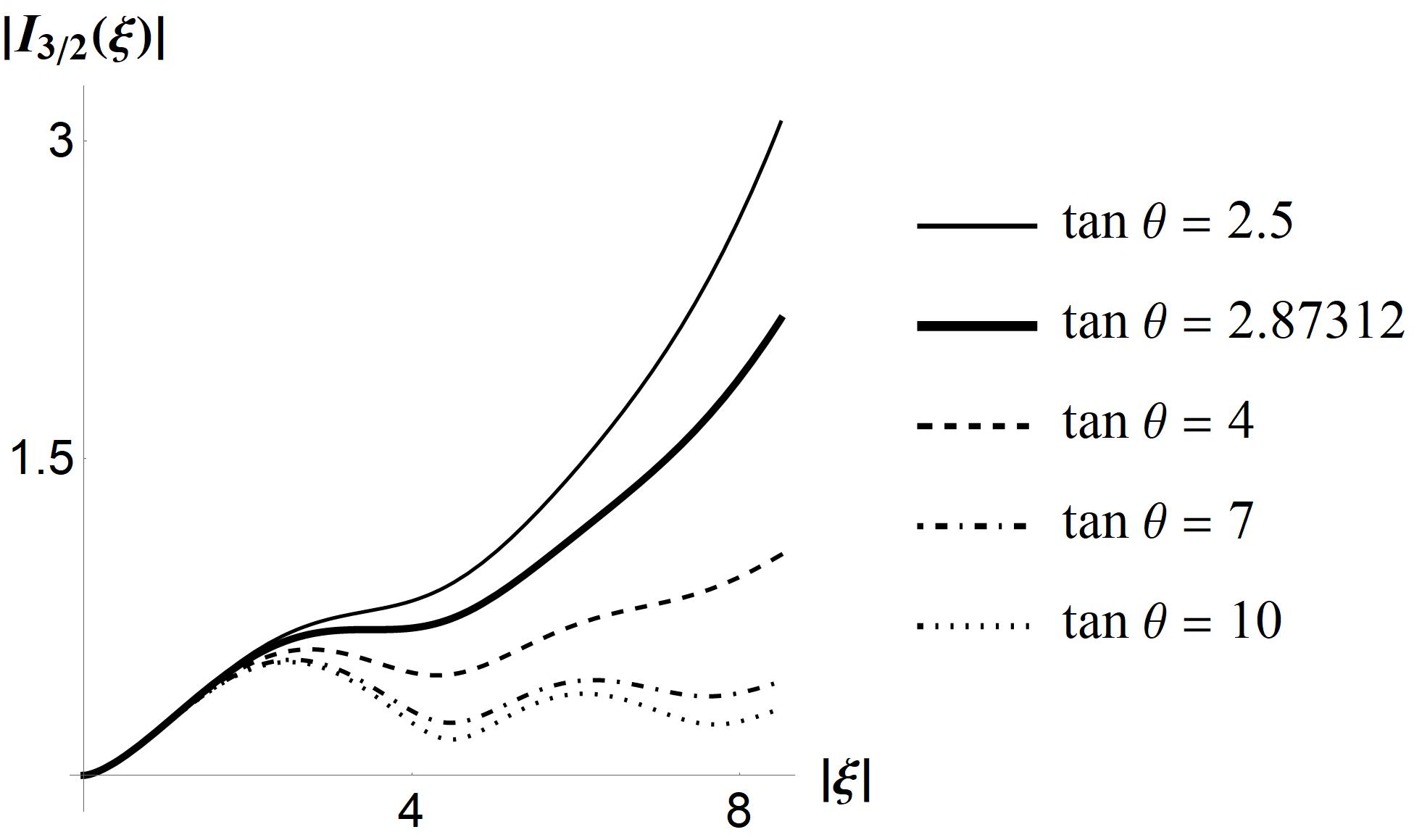}
\end{center}
\caption{The dependence of the function $|I_{3/2}(\xi)|$ on $|\xi|$ for several values of the angle $\theta$. The thick line is plotted for the approximate threshold value of $\theta$.}
\label{I32}
\end{figure}

\begin{figure}[h]
\begin{center}
\includegraphics[width=0.50\textwidth,angle=0]{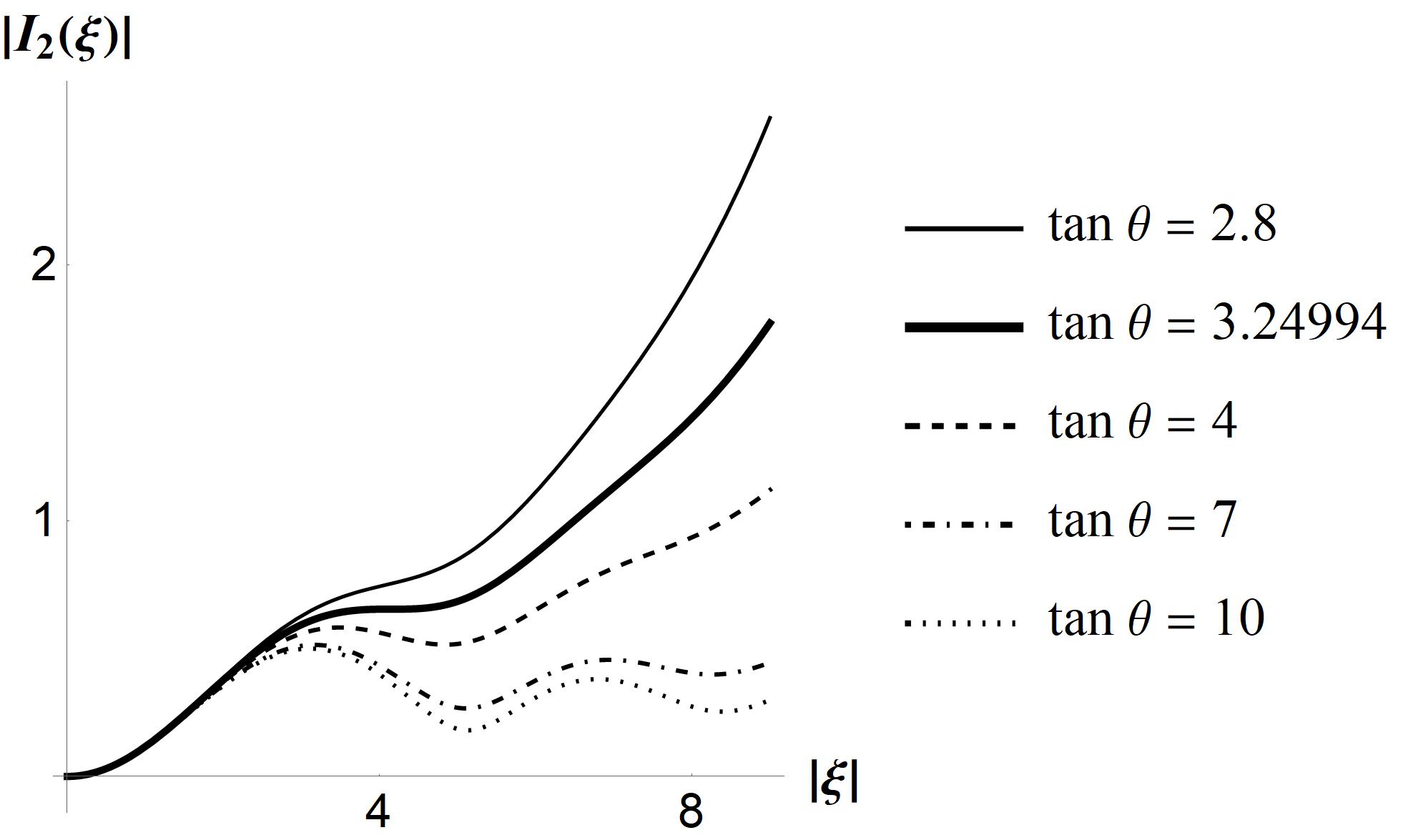}
\end{center}
\caption{Same as in Fig.~\ref{I32} but for $|I_{2}(\xi)|$.}
\label{I2}
\end{figure}

One can say that the Bessel functions $I_\nu(\xi)$ of certain complex argument $\xi$ interpolate between Bessel functions of the first kind (imaginary argument) and hyperbolic ones (real argument). This implies that the beams in question exhibit to some extent an interesting structure regarding the spatial distribution of the radiation intensity. Substituting $\xi$ in the polar form, i.e., $\xi=|\xi|e^{i\theta}$, the dependence of the absolute value of $I_\nu(\xi)$ on $|\xi|$ for different values of the parameter $\theta$ can be drawn. This is presented in Figures~\ref{I32} and~\ref{I2} for two exemplary values of the index $\nu$ ($\nu=3/2$ and $\nu=2$).

These two plots do not reveal any essential qualitative differences between half-integer and integer indices of the Bessel function and are, to a certain extent, typical for arbitrary values $\nu=n/2$.

\begin{figure}[b]
\begin{center}
\includegraphics[width=0.5\textwidth,angle=0]{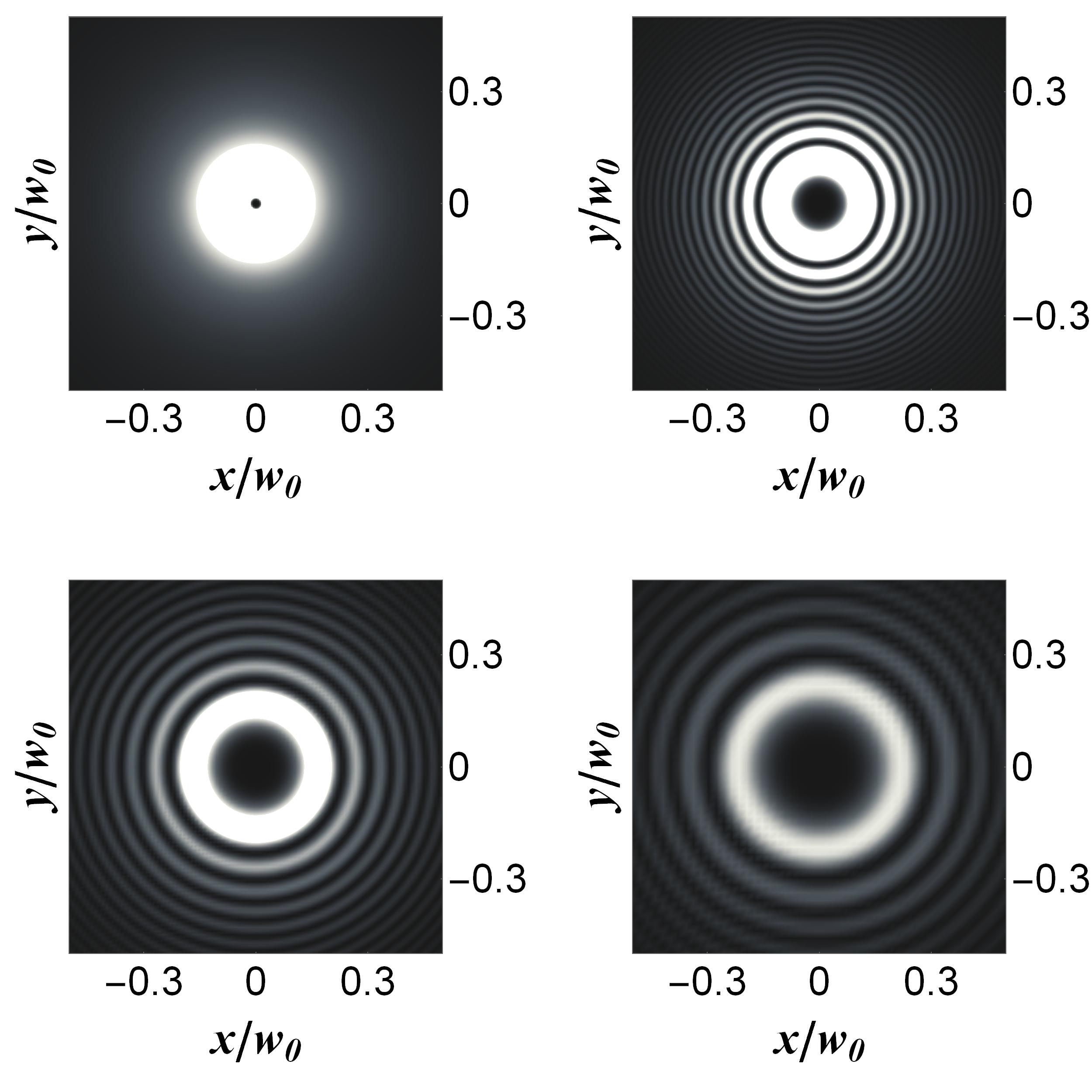}
\end{center}
\caption{Rings of high intensity (bright circles) and low intensity (dark circles) of a SHBG beam for $n=3$ in four perpendicular planes $z=2\cdot 10^{-4}z_R, 3\cdot 10^{-3}z_R, 6\cdot 10^{-3}z_R, 10^{-2}z_R$. The value of $\gamma$ is fixed to be $0.9999$. For lower values of $\gamma$ the effect appears for higher $z$, where the energy density of the wave weakens.}
\label{rings}
\end{figure}

A generic behavior observed in the diagrams is the occurrence of a series of maxima and minima when a certain threshold value of the argument $\theta$ of the complex $\xi$ is exceeded. The approximate values of these thresholds for the plotted functions are indicated in the figures. These maxima and minima develop into dark and bright rings representing areas of low and high light intensity. Looking at the form of the composite argument of the function $I_{n/2}$ as given in Eq.~(\ref{qhbg}):
\begin{equation}
\frac{\chi\, r^2}{\alpha^2(z)-\chi^2}=\frac{r^2}{w_{0}^2}\cdot\frac{\gamma}{1-\gamma-(1+\gamma)z^2/z_R^2-2iz/z_R},
\label{abe}
\end{equation}
one sees, that 
\begin{equation}
\tan\theta=\frac{2}{n+1}\varUpsilon(z)=\frac{2z/z_R}{1-\gamma-(1+\gamma)z^2/z_R^2}.
\label{tte}
\end{equation}
For $0\leq z < z_R \sqrt{\frac{1-\gamma}{1+\gamma}}$ the value of $\tan\theta$ increases from $0$ to $\infty$ thereby exceeding at a certain point the threshold value. This means that for 
\begin{equation}
z>z_{\mathrm{crit}}=z_R\, \frac{\sqrt{1+(1-\gamma^2)\tan^2\theta_n}-1}{(1+\gamma)\tan\theta_n},
\label{ztr}
\end{equation}
where $\theta_n$ denotes the threshold angle (dependent on the Bessel index $n$), initially one central ring splits into a number of concentric rings surrounding the vortex line. For $\gamma$ approaching unity this critical value of $z$ becomes relatively minor and can be estimated to be 
\begin{equation}
\frac{z_{\mathrm{crit}}}{z_R}\approx\frac{1}{2}\,(1-\gamma)\tan\theta_n.
\label{zcrit}
\end{equation}
Considering $\gamma$ values close to $1$ has that advantage that the splitting of the main ring into daughter ones occurs close to the beam's spot, i.e., where the intensity of the wave has not yet been diluted by diffraction.

Rings which emerge and vanish are not connected (as it is e.g. in the case of the Bessel beam) with zeros of the Bessel function, but only with minima and maxima, as it is presented in Figures~\ref{I32} and~\ref{I2}. Therefore,  they do not seem to be subject to any conservation law.

As can be seen from the positions of the maxima and minima in Figures~\ref{I32} and~\ref{I2}, a further shift along the beam axis is accompanied by an expansion of the rings, as visualized in Fig.~\ref{rings} for $n=3$. Other values of $n$ (even or odd ones) reveal the same effect. The external rings become thiner and thiner due to the presence of $r^2$ in~(\ref{abe}), and shallower because of the Gaussian damping factor as well as the behavior of $|I_\nu|$ plotted in the figures.

\begin{figure}[t]
\begin{center}
\includegraphics[width=0.35\textwidth,angle=0]{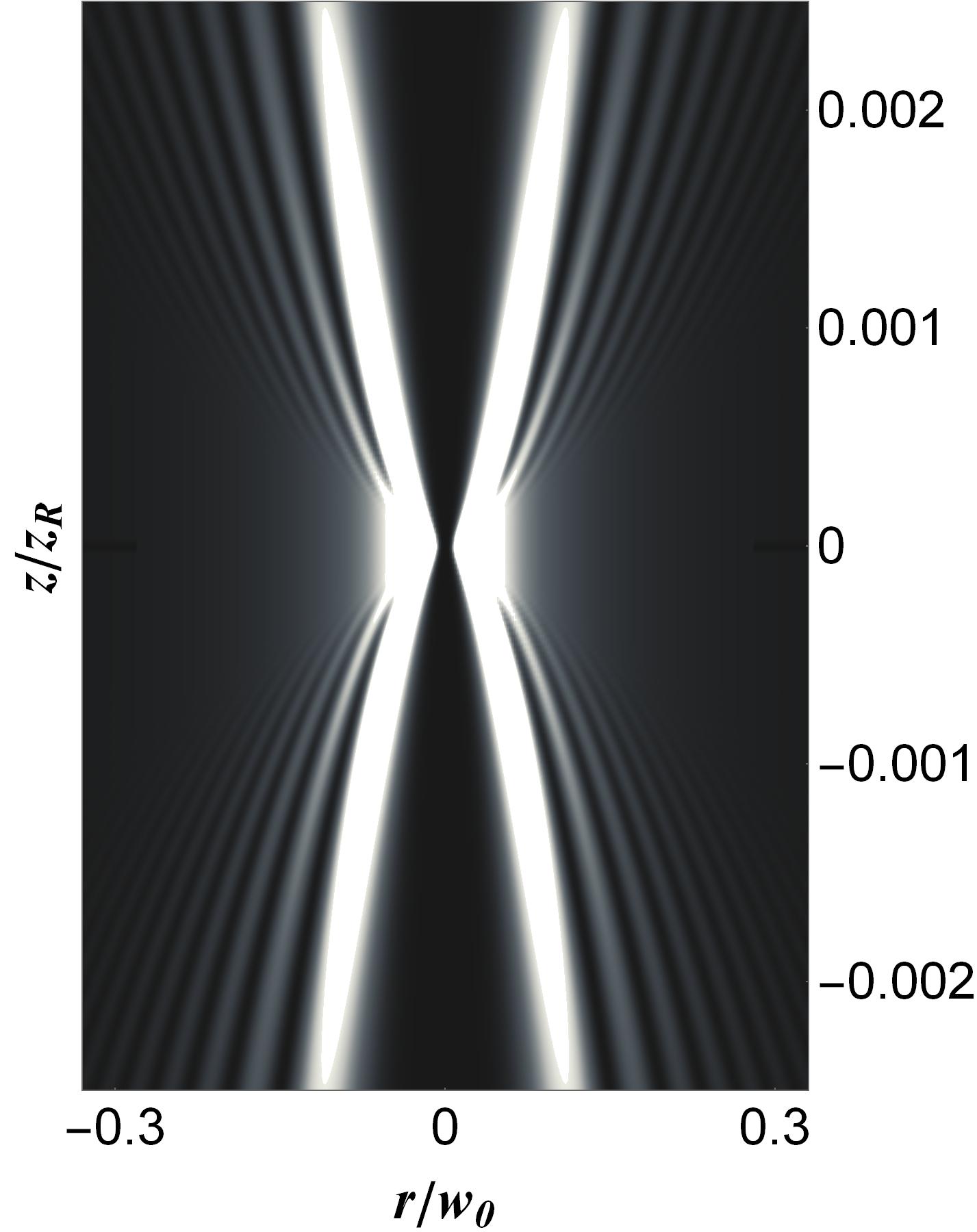}
\end{center}
\caption{Splitting of the region of high irradiance into a series of concentric rings observed in the projection onto the axial plane. The values of parameters are identical as in Fig.~\ref{rings}. For other values of the index $n$ the effect occurs identically.}
\label{sepring3}
\end{figure}

The quantity $z_{\mathrm{crit}}/z_R$ as defined by the formula~(\ref{ztr}) represents a decreasing function of the parameter $\gamma$. Therefore, for declining values of $\gamma$ the separation of annuli occurs for increasingly larger quotient $z/z_R$. Due to the presence of the factor $\gamma$ in the numerator of~(\ref{abe}), such a structure manifests itself, however, for growing values of $r$, where the Gaussian factor tends to dampen minima and maxima, making them of little eventual practical use. For instance for $\gamma\approx 0.1$, it can be estimated from equations~(\ref{psiapp}) and~(\ref{abe}) that the Gaussian factor yields an extra factor of $10^{-5}$ at the point where the first ring emerges. 
This is even more pronounced for $\gamma\rightarrow 0$ when, as we already know, the SHBG beam gets converted into ordinary Gaussian beam which does not exhibit any annular structure at all. Formally $z_{\mathrm{crit}}$ does still exist as equations~(\ref{tte}) and~(\ref{ztr}) say: 
\begin{equation}
\frac{z_{\mathrm{crit}}}{z_R}\longrightarrow \tan\frac{\theta_n}{2}
\label{zzca}
\end{equation}
but new annuli only emerge (if at all) in the very remote zone (for $r\approx w_0/\sqrt{\gamma}\rightarrow \infty$), where no longer does it make sense to talk about a beam at all.

In order to better visualize the beam's intensity distribution, it is again depicted in Fig.~\ref{sepring3} but this time in the plane containing the propagation direction $z$. The ring-merging and ring-splitting phenomenon is clearly visible. 

The following figure (Fig.~\ref{ringcomp}) presents for comparison the intensity of a standard Bessel-Gaussian beam, which also shows an annular character, but no splitting of rings or bending of annular surfaces is observed. 

\begin{figure}[h!]
\begin{center}
\includegraphics[width=0.5\textwidth,angle=0]{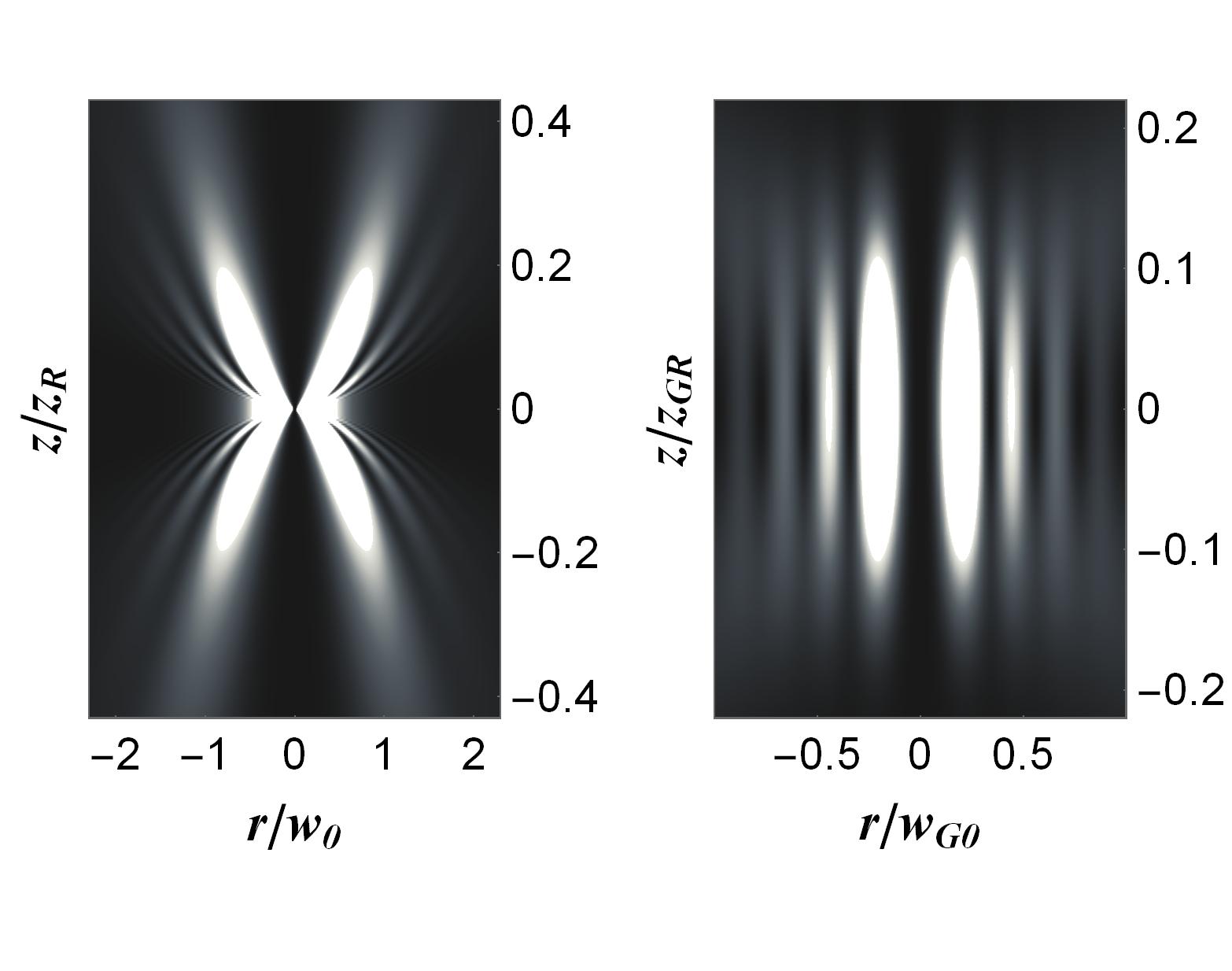}
\end{center}
\vskip -3em
\caption{Comparison of the irradiance for an SHBG beam (left diagram) and a regular Bessel-Gaussian beam (right diagram) in the plane containing the wave propagation axis. In both cases $n=2$.}
\label{ringcomp}
\end{figure}

\section{Phase distribution}

In the present section the beam's phase is analyzed. The overall phase factor, which can be absorbed into the normalization constant $N_n$, as well as time-dependent one are not accounted for.  Moreover, in order to make the evolution of the phase more clear it is convenient to omit the term $-ikz$ in the exponent of~(\ref{qhbg}). In conjunction with the expression $in\varphi$, it is responsible for the continuous rotation as the wave propagates along the $z$-axis, thereby endowing it with a vortex character. Dropping this term in our analysis and in the figures corresponds {\em de facto} to considering lines of constant phase in the $xy$ plane, which spins so as to compensate for the rotation of the vortex and, as a result, allows to better visualize the behavior of the phase. 

\begin{figure}[h]
\begin{center}
\includegraphics[width=0.50\textwidth,angle=0]{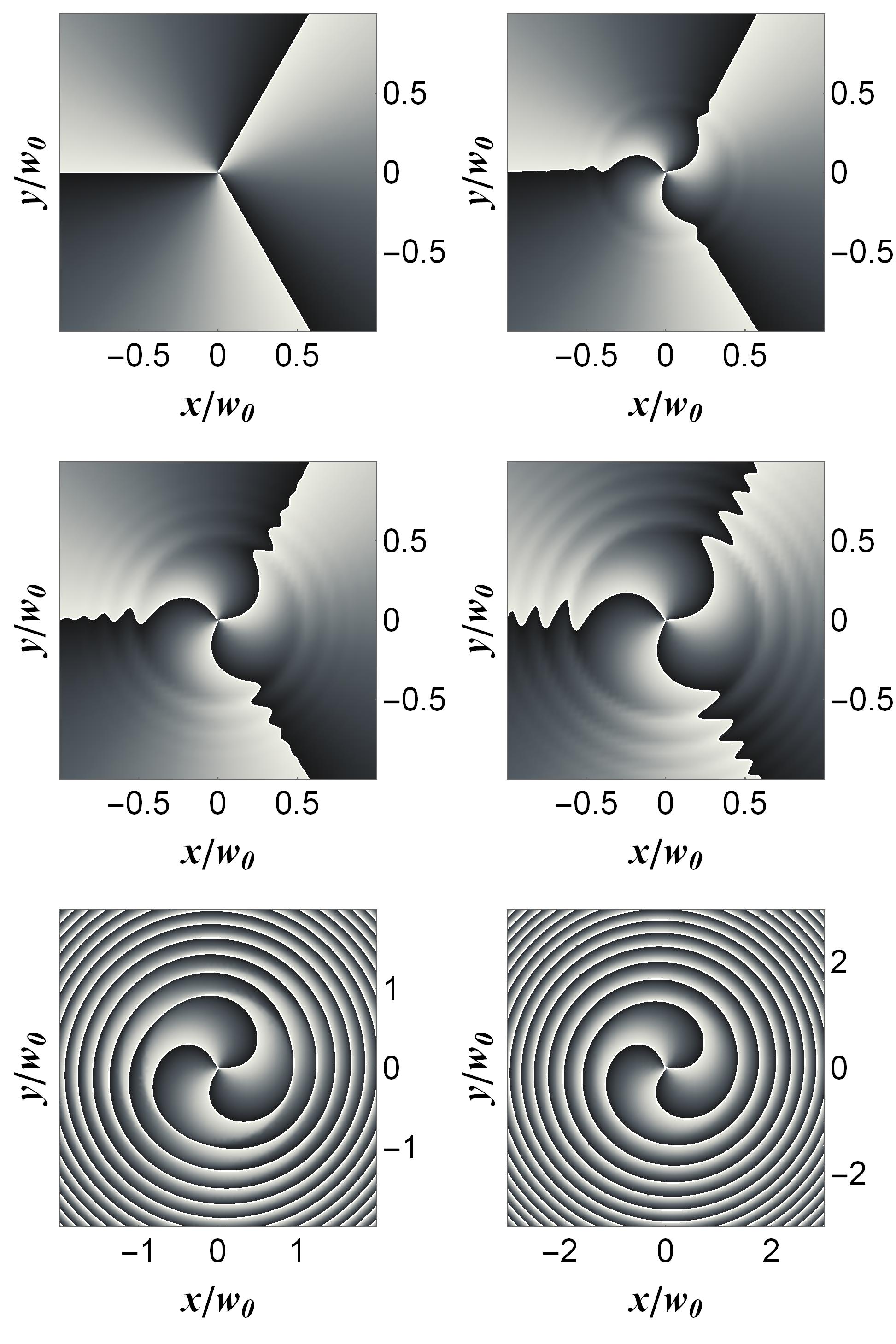}
\end{center}
\vskip -1em
\caption{The phases of the SHBG beam depicted in six planes: $z/z_R = 0, 0.02,0.03,0.04,0.1,0.2$. The value of the phase, modulo $2\pi$, is represented continuously by means of the grayscale from $-\pi$ (black color) to $\pi$ (white color). The spiral on the two last plots is typical for all the Gaussian-type beams because of  the interplay between phase factors $e^{in\varphi}$ and $e^{ikr^2/R(z)}$. The additional rotation of the entire picture with increasing $z$ appears if the factor $e^{-ikz}$ is taken into account.}  
\label{Phases13}
\end{figure}

For small values of $r$, with the use of the property~(\ref{lims}), the lines of fixed phase in the perpendicular rotating plane have the form
\begin{equation}
n\varphi+\frac{n+1}{2} \arctan \frac{2z/z_R}{1-\gamma-(1+\gamma)z^2/z_R^2}={\cal C},
\label{ph1}
\end{equation}
with ${\cal C}$ denoting a constant. Let us choose for instance ${\cal C}=\pm\pi,\pm 3\pi, \ldots$.
Since the above expression is independent on $r$ it represents $n$ straight uniformly spaced radial lines (still in the area close to the $z$-axis) rotated with respect to one another by the angle of $2\pi/n$. If one chooses $n=3$, and for $z=0$, these angles obviously are $\varphi=\pi/3, \pi, 5\pi/3$. As the wave propagates along the $z$-axis the last term is no longer vanishing and contributes to the initial (by ``initial'' we mean ``for small $r$'') value of $\varphi$. This term is an increasing function of $z$, at least for $z\ll z_R$, and therefore the layout of the lines in the middle is turned clockwise in the subsequent plots of Fig.~\ref{Phases13} by a certain angle to be denoted with $\Delta\varphi(z)$.

As $r$ increases these lines deviate from the initial value of $\varphi$, undergo some oscillations originating from the properties of the Bessel function, and finally assume asymptotically the form of a straight lines again, twisted with respect to the initial ones by the angle $-\Delta\varphi(z)$. This is clearly visible in the first four plots of Fig.~\ref{Phases13}. It is so since, with the use of~(\ref{psiapp}), the lines of the constant phase are now determined by the condition
\begin{equation}
n\varphi+\frac{r^2}{w_0^2}\cdot \frac{z/z_R}{1+z^2/z_R^2}={\cal C},
\label{ph2}
\end{equation}
and for $z\ll z_R$ the second term may be neglected (at least for moderate values of $r$, which are of interest). However, with increasing value of $z$, the second term of~(\ref{ph2}) is no longer negligible (it achieves its maximum value for $z=z_R$) and reveals the spiral character due to the presence of $r^2$, as demonstrated on the last two graphs. This is the typical behavior for Gaussian beams.

\begin{figure}[h]
\begin{center}
\includegraphics[width=0.50\textwidth,angle=0]{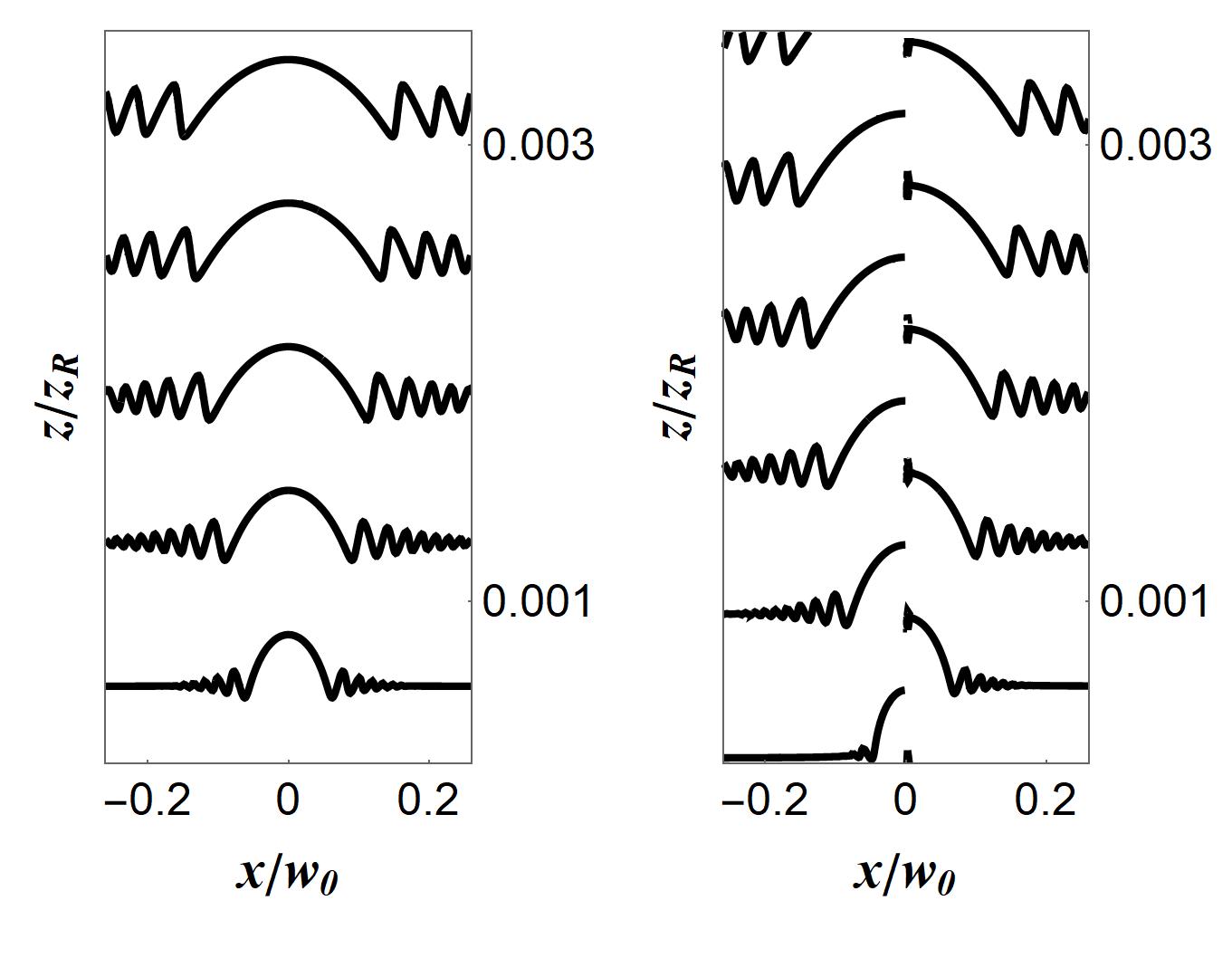}
\end{center}
\caption{The intersection of wavefronts of SHBG beams with the axial plane $xz$ for $n=2$ (left plot) and $n=3$ (right plot). The wavefront surfaces correspond are drawn for integer values of $\pi$.}
\label{Phases3}
\end{figure}

Fig.~\ref{Phases3} illustrates the intersection of the surfaces ``$\mathrm{phase}=0$'' and ``$\mathrm{phase}=\pi$'' (this time with the factor $e^{-ikz}$ included) with the plane $y=0$. For even values of $n$ the rotation around the $z$-axis produces the phase factor being the multiplicity of $2\pi$, and therefore the continuous contours are observed on the first plot. In contrast, the second plot performed for odd values of $n$, shows discontinuities since the rotation about the $z$-axis produces the additional factor of $e^{i\pi}$.

\begin{figure}[b]
\begin{center}
\includegraphics[width=0.4\textwidth,angle=0]{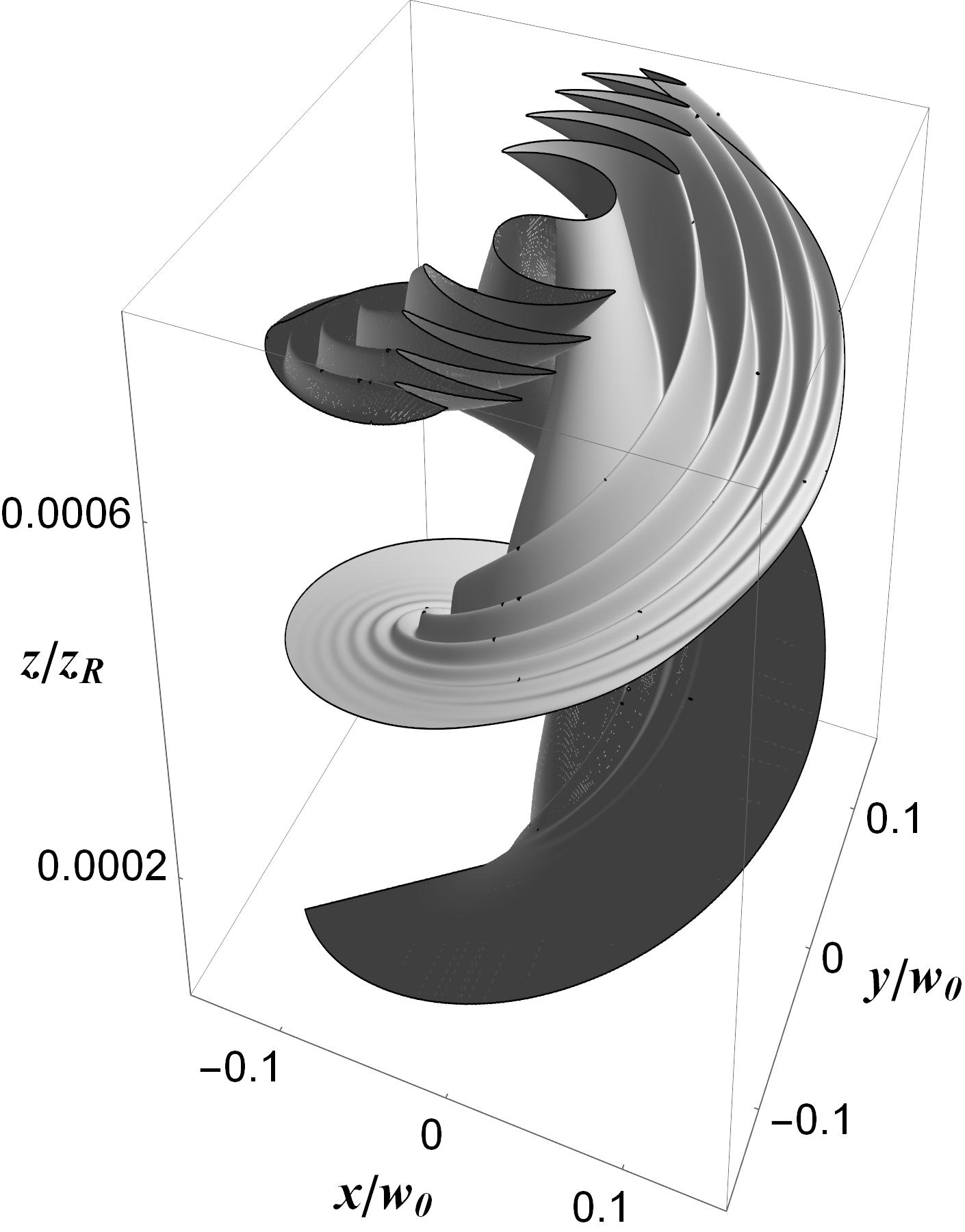}
\end{center}
\caption{Two entangled surfaces of constant phase, corresponding to the values $0$ and $\pi$ drawn for $n=1$. It is visible that the cut with the plane $y=0$ leads to discontinuities observed in Fig.~\ref{Phases3}. The undulating character of the surfaces in a radial direction is also exposed.}
\label{Phases3D1}
\end{figure}

Fig.~\ref{Phases3D1} shows a real three-dimensional representation of the surfaces of constant phase at values $0$ and $\pi$. They exhibit the helicoidal character typical of a vortex, modulated by ``oscillations'' originating from the Bessel function. It is also apparent that the discontinuities that appear in Fig.~\ref{Phases3} are solely the consequence of cutting the $3D$ surfaces with a vertical plane. This is a common feature for any beam bearing orbital angular momentum of an odd quantum number.

\section{Summary}\label{sum}

In the present work, a new family of cylindrical paraxial beams has been obtained in an analytical way, within the scalar approximation, by means of the Hankel transform. In principle they could also be found by directly solving the paraxial Helmholtz equation, but the method based on the Hankel transform is fairly universal~\cite{trhan} and quite simple. These beams, as one could it vaguely describe, ``interpolate'', during propagation along the $z$-axis, between beams specified by hyperbolic and regular Bessel functions of the first kind. Thus, during this propagation, high-irradiance rings get split (or merged). 

These analytical expressions can provide a starting point, by the construction of~\cite{ibbz}, for deriving the full vector modes, as solutions of the unapproximated Maxwell's equations. The scalar envelopes can serve then as the corresponding Whittaker potentials~\cite{whi}.

It seems that this phenomenon could be exploited in optical trapping of neutral polarizable particles based on intensity gradients, such as optical tweezers~\cite{pad}. One can imagine the segregation of particles into different channels (i.e., annular surfaces), due to their slightly different gradients, depending on the microscopic properties or initial states. The existence of an additional parameter $\gamma$ provides a theoretical possibility to control this process both as regards the spatial arrangement and the depth of potential valleys.

It would appear that the most direct method of generating beams with these characteristics is by appropriately illuminating a computer-generated hologram or computer-controlled spatial light modulator.

\end{document}